\documentstyle[prl,aps,epsfig]{revtex}

\begin{document}

\draft

\title{Optical Generation of Vortices in trapped Bose-Einstein condensates}
\author{{\L}. Dobrek$^1$, M. Gajda$1,^2$, M. Lewenstein$^3$,\\
K. Sengstock$^4$, G. Birkl$^4$,  and  W. Ertmer$^4$}
\address{$^1$College of Science, 
02668 Warsaw}
\address{$^2$Institute of Physics, 
Polish Academy of Sciences, 02668 Warsaw}
\address{$^3$Institut f\"ur Theoretische Physik, Universit\"at Hannover, 
30167 Hannover}
\address{$^4$Institut f\"ur Quantenoptik, Universit\"at Hannover,
30167 Hannover}
\date{\today}
\maketitle

\begin{abstract} 
We demonstrate numerically  
the efficient generation of vortices in  Bose-Einstein condensates (BEC)
by using a  ``phase imprinting'' method. The method consist of 
passing a far off resonant laser pulse through an absorption plate 
with azimuthally dependent
absorption coefficient, imaging the laser beam onto a BEC, and thus creating the 
corresponding non-dissipative Stark shift potential and
condensate phase shift. 
In our calculations we take into account experimental 
imperfections. We also propose an interference method to detect vortices by
coherently pushing part of the condensate using optically induced Bragg scattering.

\end{abstract}

 
\narrowtext


One of the remaining 
challenges of the physics of trapped Bose-Einstein condensates (BEC)
\cite{WiemanBEC,KetterleBEC,HuletBEC,KleppnerBEC},  
concerns the demonstration of their superfluid behavior.
Superfluidity is inevitably related to the existence of 
vortices and persistent currents in BEC, which so far 
have not been observed experimentally, despite serious efforts
\cite{priv}. Two aspects of the vortex problem have been studied intensively:
i) in rotating traps vortices appear 
in a natural way as thermodynamic ground states with 
quantized angular momentum \cite{cool}. Stability
and other properties of vortices in rotating traps have been throughly 
discussed in Ref. \cite{Strin,Rokhsar,Clark}; ii) in stationary traps
creation of vortices (or related dark solitons in 1D) requires 
the use of {\it dynamical means}, and an independent stability analysis.
 
Several methods were proposed to generate vortices in non-rotating traps:
stirring the condensate using a blue detuned laser \cite{Jackson,Ballagh},
or  several laser beams \cite{Marzlin2}; 
adiabatic passage \cite{Dum} or Raman transitions 
\cite{Marzlin1} in bi-condensate systems. Such vortices are
typically not  stable, and can exhibit dynamical or energetic instability.
In the first case vortices decay rapidly, in the second the vortices 
are stable within the framework of the mean field theory, 
and their corresponding decay requires to take into 
account interactions of the BEC with the thermal cloud. 
In the latter case the vortex dynamics is expected to  
be  sufficiently slow, and  thus experimentally accessible \cite{stabn}.

In this Communication we propose and investigate yet another procedure  
of vortex generation using ``phase imprinting''. This method consists of 
i) passing a far off resonant 
laser pulse through an absorption plate whose 
absorption coefficient depends on the rotation angle $\varphi$ around the 
propagation axis and ii) creating the corresponding
Stark shift potential inside a BEC by imaging the laser pulse onto the condensate
which leads to a $\varphi$ dependent phase shift in the condensate 
wave function.
This method is  very efficient and robust, 
and allows for engineering of a variety of excited states of BEC containing 
vortices. In the ideal case the method allows to generate 
genuine vortices with integer angular momenta. In the presence of 
imperfections, typically more complex vortex 
patterns are generated. 

We suggest an interference based method 
of vortex detection, in analogy to methods 
used in nonlinear optics\cite{nonlin}. 
In the context of matter waves similar  methods were 
proposed in references\cite{Bolda,revue}. Our idea is to combine this method with the 
recently developed techniques of Bragg diffraction for BEC 
manipulation\cite{Bill}.

Before we turn to details,
we should stress that
the dynamical generation of vortices differs from 
the case of rotating traps, in which a {\it pure vortex} state with
angular momentum $L_z=1$ (in units of $\hbar$) is selected in the process 
of reaching the 
equilibrium. In our case, generation of {\it pure} vortices a requires fine 
tuning of parameters which is hard to achieve in experiments. 
Our method is suitable for creation of generic states with {\it vorticity}
\cite{Bolda}, i.e. states with several vortex lines, around which
the circulation of velocity does not vanish 
\cite{Nozieres}.
 
Dynamics of BEC, and thus  the process 
of creation and evolution of vorticity at zero temperature is
well described by the time-dependent Gross-Pitaevskii equation for 
the wave function $\psi ({\bf r},t)$:
\begin{equation}
\label{GP}
i \hbar \partial_{t} \psi = 
\left( \frac{-\hbar^2 \nabla^2}{2M} 
+V_{NL}({\bf r},t)+ V_t({\bf r})+ 
V_l({\bf r},t) \right) \psi,
\end{equation}
where $V_t({\bf r})=M(\omega_x^2x^2+\omega_y^2y^2+ \omega_z^2z^2)/2$ 
is the external trap potential which we assume to 
be harmonic, $M$ is the mass of the atom, and $\omega_x, \omega_y,
\omega_z$ are the trap frequencies. The nonlinear 
term $V_{NL}({\bf r},t)= g |\psi({\bf r},t)|^2$ describes the 
mean field two-body repulsive interaction 
whose strength $g$ is  related to the scattering length $a$ by 
$g= 4 \pi N \hbar^2 a/M$, where $N$ is the total number of condensed 
particles. The term  
$V_l({\bf r},t)$ describes an effective potential created by an external
laser beam impinging on the condensate 
after passing through a plate with appropriately modulated absorption
coefficient. In the following we will study the two dimensional 
version of the Eq. (\ref{GP}), in which we replace $g$ by $g/D$, 
where $D$ is the characteristic depth chosen such that the ground state
chemical potentials of the 3D and 2D systems are equal.

In this Communication we show that a short pulse of light with a typical
duration of the order of fractions of  microseconds with properly modulated 
intensity profile  creates vorticity in a Bose-Einstein condensate initially 
in its ground state. If the incident light is detuned 
far from the atomic transition frequency its main effect on the atoms
is to induce a Stark shift of the internal energy levels. As the intensity
of light depends on the position, the Stark shift will also be position 
dependent as will be the phase of the condensate.

The main feature characterizing a vortex is related to the particular 
behavior of the phase of the wave-function at the vortex line: the phase
``winds up'' around this line i.e. it changes by an integer multiple 
$m$ of $2\pi$ on a path surrounding the vortex. Index $m$ is the 
vortex charge. The light beam, before impinging 
on the atomic system, is shaped by an absorption plate whose absorption
coefficients varies linearly with rotation angle $\varphi$ around the plate axis (see
Fig.~1).  As a result, $V_l({\bf r},t)$ depends 
on the distance from the propagation axis $\rho$, and the azimuthal angle
$\varphi$.

In the ideal case this absorption plate causes a real jump of the potential 
at, say $\varphi=0$. In this case we can model the potential for $0\le t \le T$ by
\begin{equation}
V_l(\rho,\varphi) =
\left\{ 
\begin{array}{ll}
\hbar I\varphi\sin^2(\pi\rho/2L) & {\rm for}\ \rho\le L \\
\hbar I\varphi &{\rm for}\ \rho > L,
\end{array}
\right.
\end{equation}
being zerofor other times. 
Here $T$ is the (square) pulse duration, whereas $I$ denotes the characteristic
Stark shift (proportional to the laser intensity); $L$ is the 
characteristic length scale, on which the absorption profile is smoothed in our
calculations in the
vicinity of the propagation--rotation axis. The use of more realistic temporal
pulse shapes does not change the results significantly. In the following we
will use square pulses, since it allows to control the pulse area in a simpler
manner.

The characteristic time and length scales for the BEC that we consider
are ms, and $\mu\mbox{m}$, respectively\cite{bounce}. Note, that for this parameter range
$T$ is short ($\simeq\mu$s), and $\hbar I$ much larger then 
other energy scales, the dynamical effect of $V_l(\rho,\varphi)$ corresponds 
approximately to a ``phase imprinting''. Since during the interaction with 
the laser, all other terms in Eq. (\ref{GP}) can be neglected, the wave 
function after switching off the pulse becomes
\begin{equation}
\psi({\bf r}, T)=\exp(-iTV_l(\rho,\varphi)/\hbar)\psi({\bf r},0).
\label{dupa}
\end{equation}
For the ideal case, choosing $IT=m$, we get the desired phase dependence
characteristic for a pure vortex state.  Unfortunately, the above picture is
oversimplified, because the wave function (\ref{dupa}) does not necessarily
vanish at the vortex line ($\rho=0$), which signifies infinite kinetic energy
and its unphysical character.  
One has to solve the full
dynamics of the system for $0\le t\le T$, taking especially into account the
kinetic energy term.  Moreover, realistic absorption plates cannot generate
the singularity in the $\varphi$--dependence of $V_l$. For this reason we smooth
the $\varphi$--dependence of the intensity $I$ on the scale of $\delta$ radians
(see Fig. 1).  ``Phase imprinting'' with this smoothed intensity distribution cannot
generate pure vortex
states --- it can, however, as we shall see below,  generate in a controlled
way states with predetermined vorticity.

The Gross-Pitaevskii equation was solved using the split operator method in
2D ($x$ and $y$ plane). The simulations were divided into two stages: 
the initial excitations stage 
of duration $T\simeq$ 0.16$\mu$s, with about 1000 time steps, and the 
second stage with a characteristic time scale of ms, and about 1000 
steps/ms. We have assumed that initially a condensate 
containing $N=100 000$ Rubidium atoms
(with $a=5.8$nm) was trapped in a disc-shaped trap of frequencies 
$\omega_x=\omega_y=2\pi\times 30$Hz, $\omega_z=2\pi\times 300$Hz. 
In this case the condensate radius was about $15\mu$m.
We have used $I=6\times 10^6$Hz, and adjusted $T$ to obtain the desired 
values of the pulse area ($IT\simeq 1,2,\ldots$). 
The $\rho$--dependence of the potential $V_l$ was smoothed with 
$L\simeq 1.7\mu$m, and the $\varphi$--dependence
with $\delta\simeq 0.15$ radians. The smallest possible value of $\delta\simeq 0.04$
radians was determined by the spatial grid size which  typically was 
512$\times$512 points in a $40\mu{\rm m}\times40\mu{\rm m}$ box. 
Additionally,  we studied the case when the laser was focused 
slightly off the trap symmetry axis. During the dynamics, after switching off 
the laser,  the following quantities are conserved:
wave function norm, mean energy, and mean $z$-component of the 
angular momentum. They were monitored in order to control the accuracy 
of numerics. The first two quantities were constant within the accuracy 
of the method, the third one in some cases exhibits slow variations 
of the order of a percent.

One of the most important questions concerning the investigation of 
vortices is an efficient
method for their detection. Experimentally monitoring density 
profiles with the necessary resolution is
difficult, since the vortex core is very small. The best 
way is to monitor the phase of the wave function in an interference 
measurement. Such interference measurements are routinely done in nonlinear 
optics\cite{nonlin}. In the context of BEC, they were first proposed 
by Bolda and Walls\cite{Bolda}, who considered the interference of 
two condensates moving towards each other. If both condensates are in the 
ground state (no vortices), one expects interference fringes as those
observed by Ketterle et al.\cite{inter}. In the case of interference of 
one condensate in the ground state with the second one in the $m=1$ vortex state,
a fork--like dislocation in the interference pattern appears. The distance 
between the interference fringes is determined by the 
relative velocity of the condensates, which can be controlled experimentally. 
This is a very efficient and clear method of vorticity detection. 
It requires, however, the use of two independent condensates.

We propose here to combine the interference method with the recently 
developed Bragg diffraction technique\cite{Bill}, which has been 
successfully used in a four wave mixing experiment\cite{4wave}. 
The idea is to transfer part of the atoms coherently to another momentum state
using one, or several stimulated two-photon Raman scattering processes. The 
procedure is the following: i) first we create the vortex, or vorticity state 
in the trap; ii) we open the trap and let the condensate expand; iii) when 
the density is reduced to values for which nonlinear interactions are 
negligible, we apply the Bragg pulses. Part of the wave function 
attains a phase factor that signifies the fact that the corresponding 
momentum was transferred to part of the atomic sample\cite{note}. The 
resulting wave function is the superposition of two vortex (or vorticity) states 
moving apart from each other, with a velocity that can be easily controlled
by choice of the angle between the Bragg beams.
In the following we will use velocities of the order of 1 mm/s, 
which allow for efficient detection  after 4-5 ms when the vortices are
about 5$\mu$m apart; iv) detection consists of optical  
imaging that is accomplished within a few $\mu$s. The interference 
patterns have a characteristic length scale of a few $\mu$m.

We shall first discuss our results for the ideal case: $\delta\le0.04$ radians,
$IT=1$, and the laser focused exactly along the trap symmetry axis. 
In this case during the first stage of the dynamics angular momentum is 
transferred to the system. The mean of  $L_z$ grows monotonically from 0 to about 1 
(in units of $\hbar$), and stays close to 1 hereafter. The dispersion $\delta L_z$
grows within the pulse duration to about $5$, returns to about 0 for $t=T$.
In the second stage of the dynamics, a hole in the density profile appears,
and the density acquires a stationary state within a fraction of a ms.  This time scale
corresponds to the inverse of the characteristic frequency, determined by the mean
field energy in the condensate, $g/\hbar V$, where $V$ is the volume of the
condensate ($V\simeq \pi 15\times 15\times 4\mu$m$^3$). It is the time required
by the second sound to travel over a distance of the order of the healing length.
The density distribution hereafter remains constant on the scale of $10-20$ms. The
dispersion  of the angular momentum, $\delta L_z$, exhibits a
stepwise growth to about $\delta L_z \simeq 5$ with fast growth periods of
duration of about  0.5 ms, followed by slow growth stages of duration of a
few ms. We observed a slight movement of the vortex off the center of the trap.

In Fig. 2a we present a typical simulation result obtained a few ms after
the creation of a vortex using the detection method of\cite{Bolda} based on the 
interference of a condensate in the vortex state with a condensate in the 
plane wave state. The characteristic fork--like pattern reflects the fact that the phase
winds up by $2\pi$ as one circulates around the vortex line. Fig. 2b shows 
the simulation of the interference pattern for the method of detection   
proposed by us. A few milliseconds after applying the 
Bragg pulses we can clearly see a double fork structure, as one should expect 
for this case. The forks are oriented in opposite directions, because the condensates
have the same helicity, but opposite velocities.  

It is interesting to compare this ideal result with more realistic simulations
obtained for $\delta=0.15$, keeping the laser focused at the center of the 
trap and the pulse area $IT=1$. In this case one excites a vorticity 
state 
which is not pure. The mean $L_z$ after the laser pulse is close to zero,
as it should be in the absence of the phase discontinuity. 
The dispersion  $\delta L_z$ behaves similarly as before, although it grows 
a little faster. 
The density profile exhibits  now the hole in the center but also a number of 
deep minima, or holes at $\varphi=0$ in the outer regions of the condensate. 
While the density distribution is difficult to interpret, the interference method allows to 
interpret them very clearly. Application of the plane wave method indicates two 
vortices of opposite winding number (topological charge), 
and perhaps two more close to the edge of the condensate (Fig. 3a).  
This basic structure is doubled, but nevertheless clearly recognizable 
using interference combined with the  Bragg scattering (Fig. 3b).

Finally, in Fig. 4 we present numerical results that mimic other possible 
experimental imperfection: offset of the laser focus by 1.7$\mu$m 
off the center of the trap (vertical in the figure plane), 
combined with the smoothed 
$\varphi$--dependence of absorption profile ($\delta=0.15$). 
In this case the behavior of $L_z$ and $\delta L_z$ is similar to 
the case presented in Fig. 3. The interference patterns, obviously, 
are more noisy and irregular. Still  one can clearly recognize  
fork--like dislocations corresponding to vortices using the plane wave interference
method (Fig. 4a), and the method proposed here (Fig. 4b).  

In our numerical simulations we have also studied situations when the 
area of the phase imprinting pulse is large, $IT > 1$. Typically  
a number of singly charged vortices aligned in geometric patterns
similar to those obtained in rotating traps\cite{Rokhsar} are created.

Summarizing, we have proposed a method of vortex generation in 
trapped Bose-Einstein condensates. The method employs optical potentials
induced by passing a laser beam through an absorption plate 
with an absorption coefficient that depends on the azimuthal angle $\varphi$.
We have also proposed a method of detecting vortices and vorticity by combining
the known interference method with the recently developed Bragg scattering 
techniques to coherently transfer atoms into a selected momentum state.

We acknowledge fruitful discussions with K. Bongs, S. Burger, 
M. Brewczyk, J.I. Cirac, S. Dettmer, W. Kr\'olikowski, and K. Rz\c a\.zewski.
This work is supported by the SFB 407 of the {\it Deutsche 
Forschungsgemeinschaft}. MG is partially supported by the KBN grant 
2 P03B 13015.


\begin{figure}
   \begin{center}
   \epsfxsize 3.5cm
   \epsffile{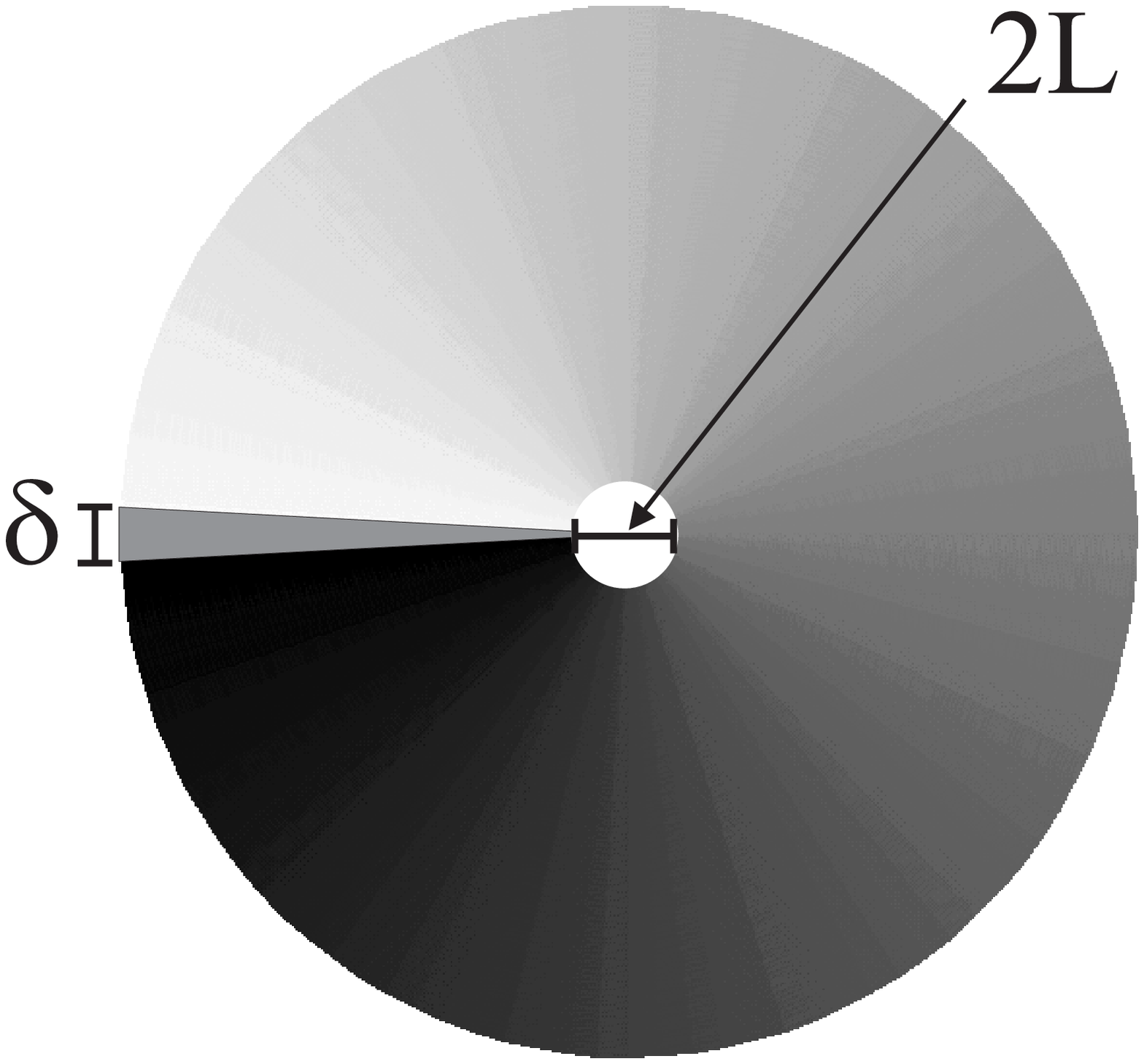}
   \end{center}
   \caption{Absorption plate. \mbox{$L$}  is the radial extension on which the absorption
    profile was smoothed in our simulations, whereas  
    \mbox{$\delta$}  is the angular extension of the smoothing.}
   \label{fig1}
\end{figure}

\begin{figure}
   \begin{center}
   \epsfxsize 6.5cm
   \epsffile{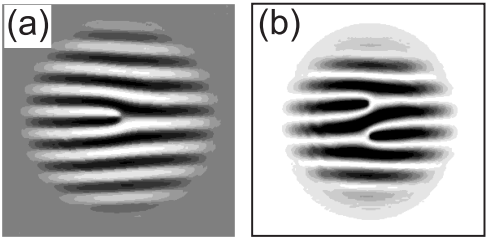}
   \end{center}
   \caption{Interference pattern of the vortex state 
           (5 ms after phase imprinting): 
           (a) superimposed  with a plane wave of 
           $k=1.8\,\mu \mbox{m}^{-1}$ moving vertically
           in the plane of the figure; 
           (b) 5 ms after applying Bragg
           pulses transferring momentum to a part of the condesate 
           (along vertical axis in the plane of the figure). 
	   The ideal case of equal splitting 
           of the whole condensate has been assumed.
           The incident pulse was
           focused at the trap center and had a sharp step 
           in the intensity profile ($\delta=0.04$).}
   \label{fig2}
\end{figure}

\begin{figure}
   \begin{center}
   \epsfxsize 6.5cm
   \epsffile{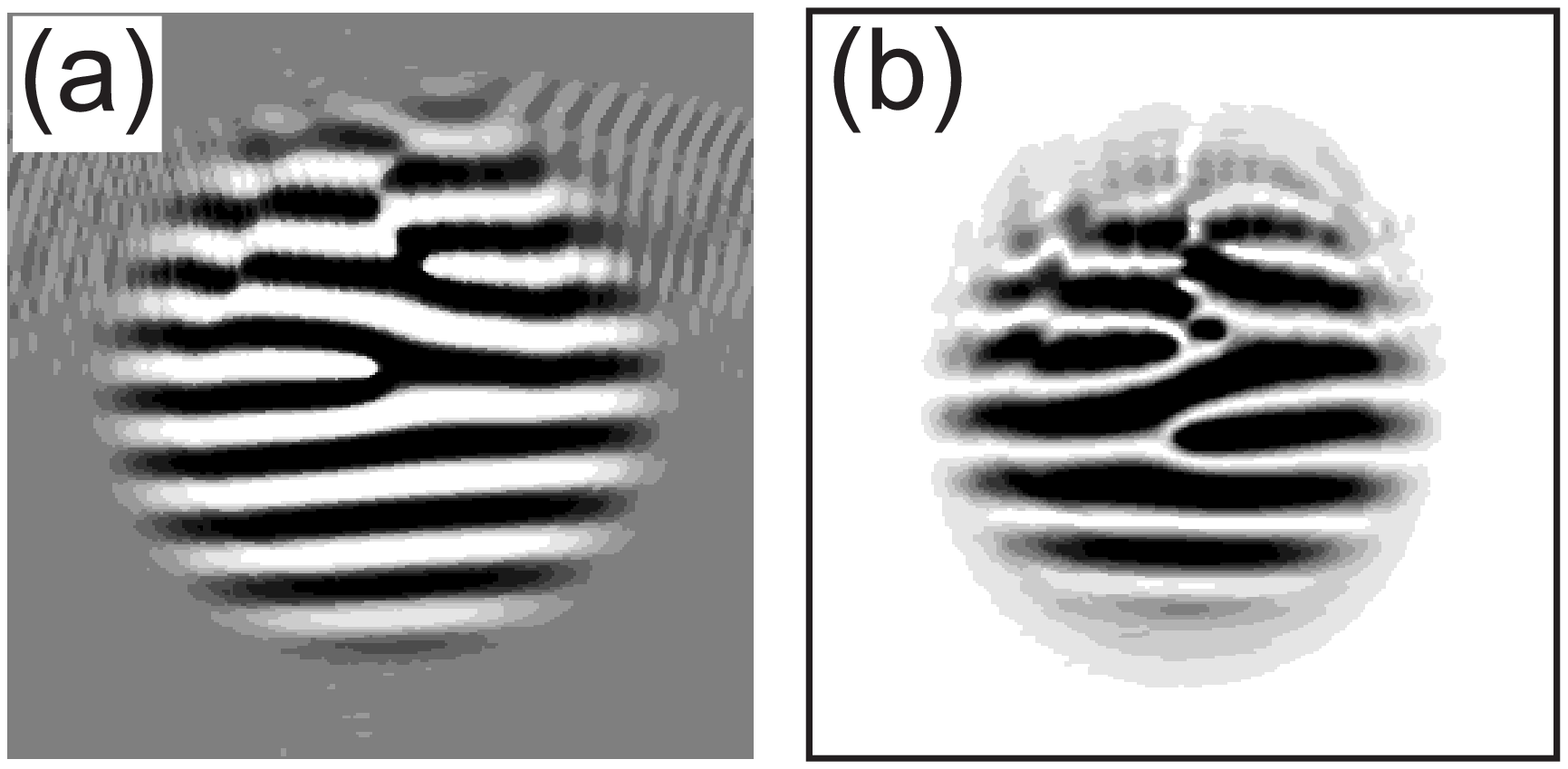}
   \end{center}
   \caption{Interference pattern of the vorticity state 
           (6 ms after the phase imprinting):
            (a) after superimposing with a plane wave as in Fig.~1a; 
            (b) after applying the Bragg pulses as in Fig.~1b.
            The incident pulse was focused at the trap center and 
		its intensity profile was smoothed over an angle $\delta=0.15$.}
   \label{fig3} 
\end{figure}

\begin{figure}
   \begin{center}
   \epsfxsize 6.5cm
   \epsffile{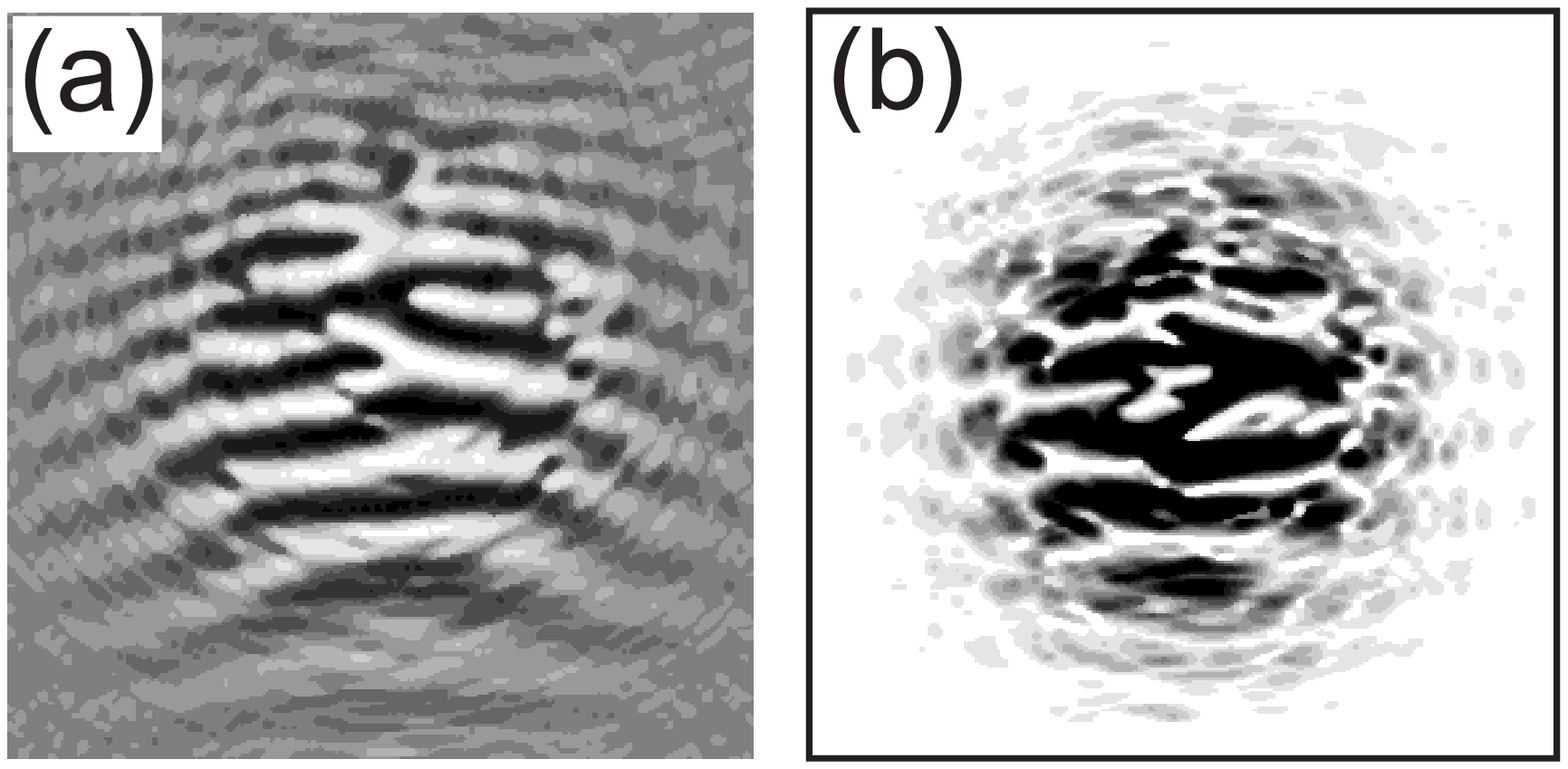}
   \end{center}
   \caption{Interference pattern of the vorticity state
            (8.5 ms after phase imprinting):
            (a) after superimposing with a plane wave as in Fig.~1a; 
            (b) after applying the Bragg pulses as in Fig.~1b.
            The incident pulse was focused $1.7\,\mu\mbox{m}$ off the 
            trap center and its intensity profile was smoothed over an angle 
            $\delta=0.15$.}
   \label{fig4}
\end{figure}

%


\begin{references}
\bibitem{note} If the resonant width of the Raman transition 
is narrower than the wave function spreading in the momentum space 
not all momentum components can experience the shift. In that case the contrast
of interference fringes will be reduced. One can however correct for this by
adiabatically opening the trap before applying Bragg pulses.
\end{references}
\end{document}